%% file: M0416_SLletter_arXiv.tex
\documentclass[useAMS,usenatbib]{mn2e}

\usepackage{graphicx}
\usepackage{txfonts}
\usepackage{amssymb,alltt}
\usepackage{color}
\usepackage{hyperref}
\usepackage{supertabular}

\hypersetup{colorlinks, linkcolor=blue}

\title[HFF Strong-Lensing Analysis of MACSJ0416.1-2403]{\textit{Hubble Frontier Fields}: A High-Precision Strong-Lensing Analysis of Galaxy Cluster MACSJ0416.1-2403 Using $\sim$200 Multiple Images }
\author[Jauzac et al. 2014]
{M. Jauzac,$^{1,2}$\thanks{E-mail:
mathilde.jauzac@dur.ac.uk} B. Cl\'ement,$^{3}$ M. Limousin,$^{4,5}$
J. Richard,$^{6}$ E. Jullo,$^4$ H. Ebeling,$^{7}$ H. Atek,$^{8}$ 
\newauthor
J.-P. Kneib,$^{8,4}$ K. Knowles,$^{2}$ P. Natarajan,$^{9}$ D. Eckert,$^{10}$ E. Egami,$^{3}$
R. Massey,$^{1}$ M. Rexroth$^{8}$  
\\
\\
$^{1}$Institute for Computational Cosmology, Durham University, South Road, Durham DH1 3LE, U.K.\\
$^{2}$Astrophysics and Cosmology Research Unit, School of Mathematical Sciences, University of KwaZulu-Natal, Durban 4041, South Africa\\
$^{3}$Steward Observatory, University of Arizona, 933 North Cherry Avenue, Tucson, AZ, 85721, USA \\
$^{4}$Laboratoire d'Astrophysique de Marseille - LAM, Universit\'e d'Aix-Marseille $\&$ CNRS, UMR7326, 38 rue F. Joliot-Curie, 13388 Marseille Cedex 13, France\\
$^{5}$Dark Cosmology Centre, Niels Bohr Institute, University of Copenhagen, Juliane Maries Vej 30, DK-2100 Copenhagen, Denmark\\
$^{6}$CRAL, Observatoire de Lyon, Universit\'e Lyon 1, 9 Avenue Ch. Andr\'e, 69561 Saint Genis Laval Cedex, France\\
$^{7}$Institute for Astronomy, University of Hawaii, 2680 Woodlawn Drive, Honolulu, Hawaii 96822, USA\\
$^{8}$Laboratoire d'Astrophysique, Ecole Polytechnique F\'ed\'erale de Lausanne (EPFL), Observatoire de Sauverny, CH-1290 Versoix, Switzerland\\
$^{9}$Department of Astronomy, Yale University, 260 Whitney Avenue, New Haven, CT 06511, USA\\
$^{10}$Astronomy Department, University of Geneva, 16 ch. d'Ecogia, CH-1290 Versoix, Switzerland}

\begin{document}

\date{Accepted 2014. Received 2014; in original form 2014 May 15}

\pagerange{\pageref{firstpage}--\pageref{lastpage}} \pubyear{2014}

\maketitle

\label{firstpage}

\begin{abstract}
\input{abstract_submitted}
\end{abstract}



\section{Introduction}
\label{intro}
\input{introduction_submitted}

\section{\textit{Hubble Frontier Fields} Observations}
\label{observations}
\input{HFFdata_submitted}

\section{Strong Lensing Analysis}
\label{SLanalysis}
\begin{figure*}
\begin{center}
\includegraphics[width=\textwidth]{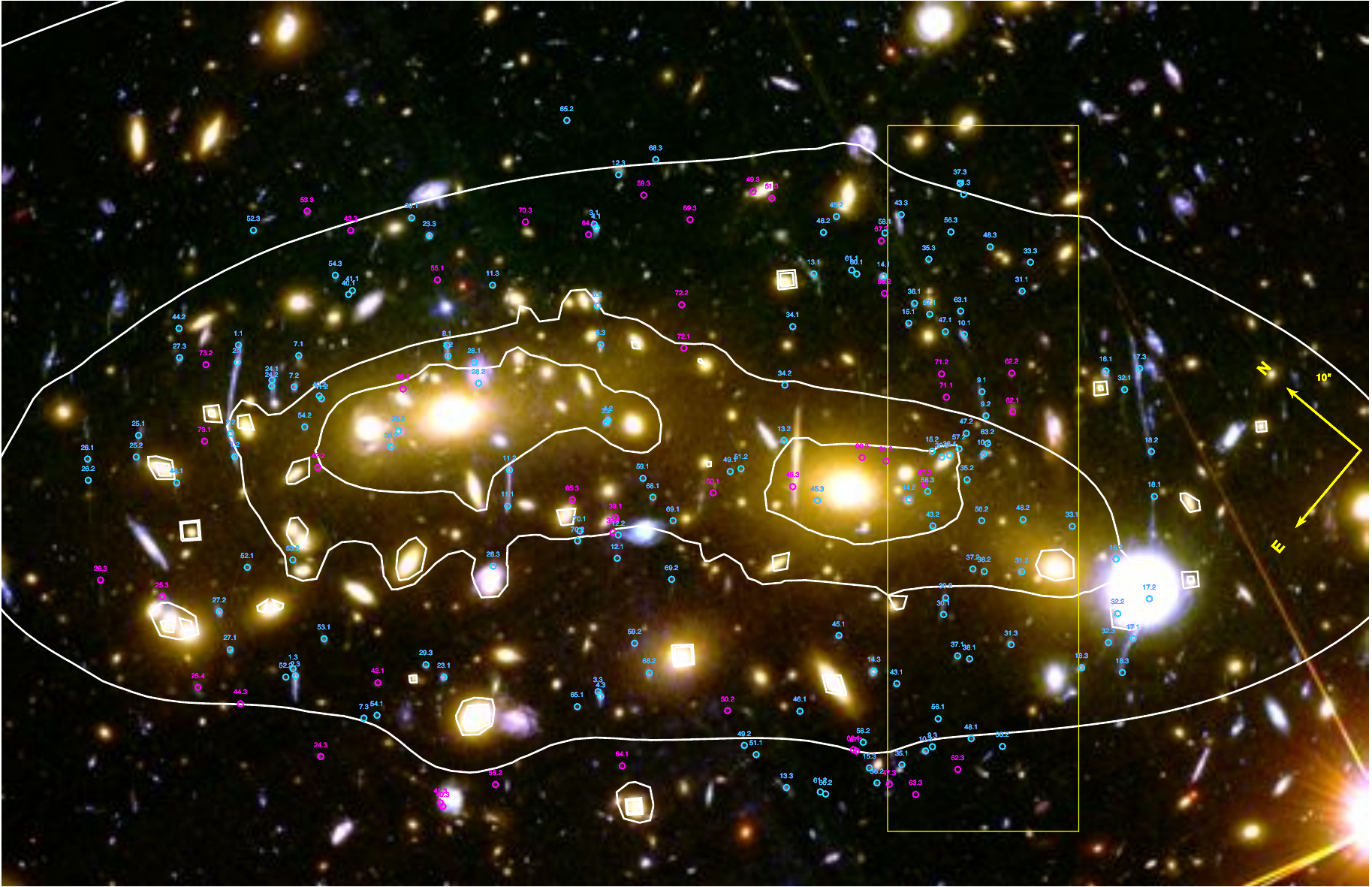}\\
\includegraphics[width=\textwidth]{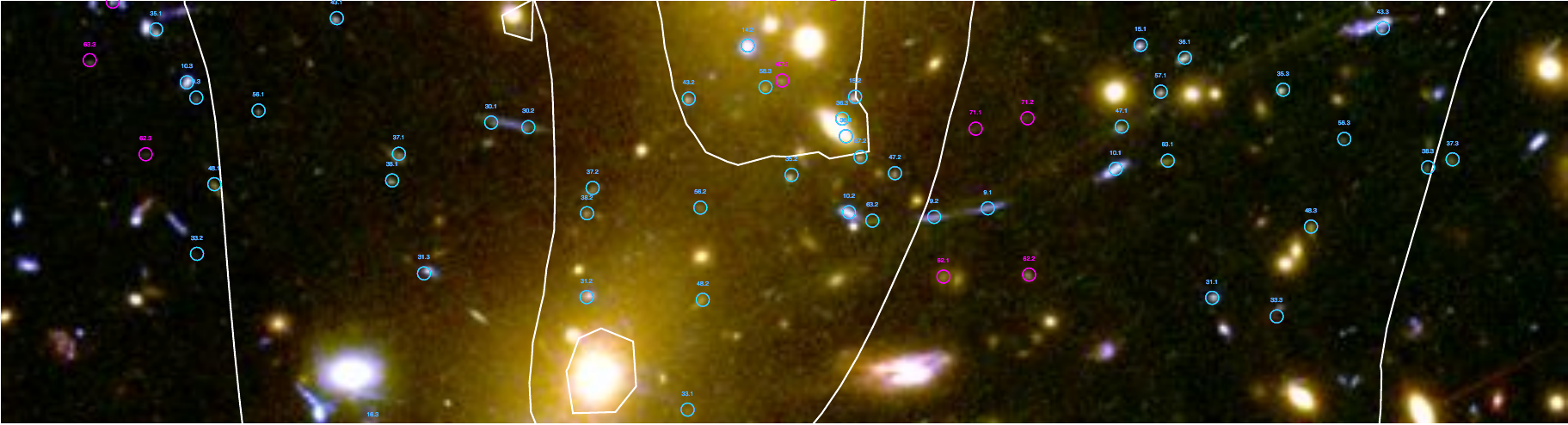}\\
\caption{Overview of all multiple-image systems used in this study. The most secure identifications, used to optimise the lens model in the \textit{image plane} (149 images) are shown in cyan; the less secure candidates (45 images) are shown in magenta. The underlying colour image is a composite created from HST/ACS images in the F814W, F606W, and F435W passbands. Mass contours of the best-fit strong-lensing model are shown in white. The yellow rectangle in the top panel highlights the zoomed region shown in the bottom panel.}
\label{multiples}
\end{center}
\end{figure*}

By carefully inspecting the deep HFF images of MACSJ0416 we identify 51 new multiple-image systems, three times as many as previously known, bringing the total to 68 multiple-image families composed of 194 individual images (see Fig.~\ref{multiples} and Table~\ref{multipletable}). Spectroscopic redshifts are presently available for nine of these systems: although constituting but a small fraction of the total, these are sufficient to calibrate the absolute mass of the lens. We are thus now able to dramatically improve the strong-lensing model of this cluster.

\subsection{Methodology}
\input{SLmethod_submitted}

\subsection{Multiple-Image Systems}
\input{SLconstraints_submitted}

\section{Strong-Lensing Mass Measurement}
\label{SLmass}
\input{SLmass_submitted}

\input{modele_SL_submitted.tex}

\section{Discussion}
\label{discussion}

\begin{center}
\begin{figure*}
\includegraphics[width=5.1cm,angle=0.0]{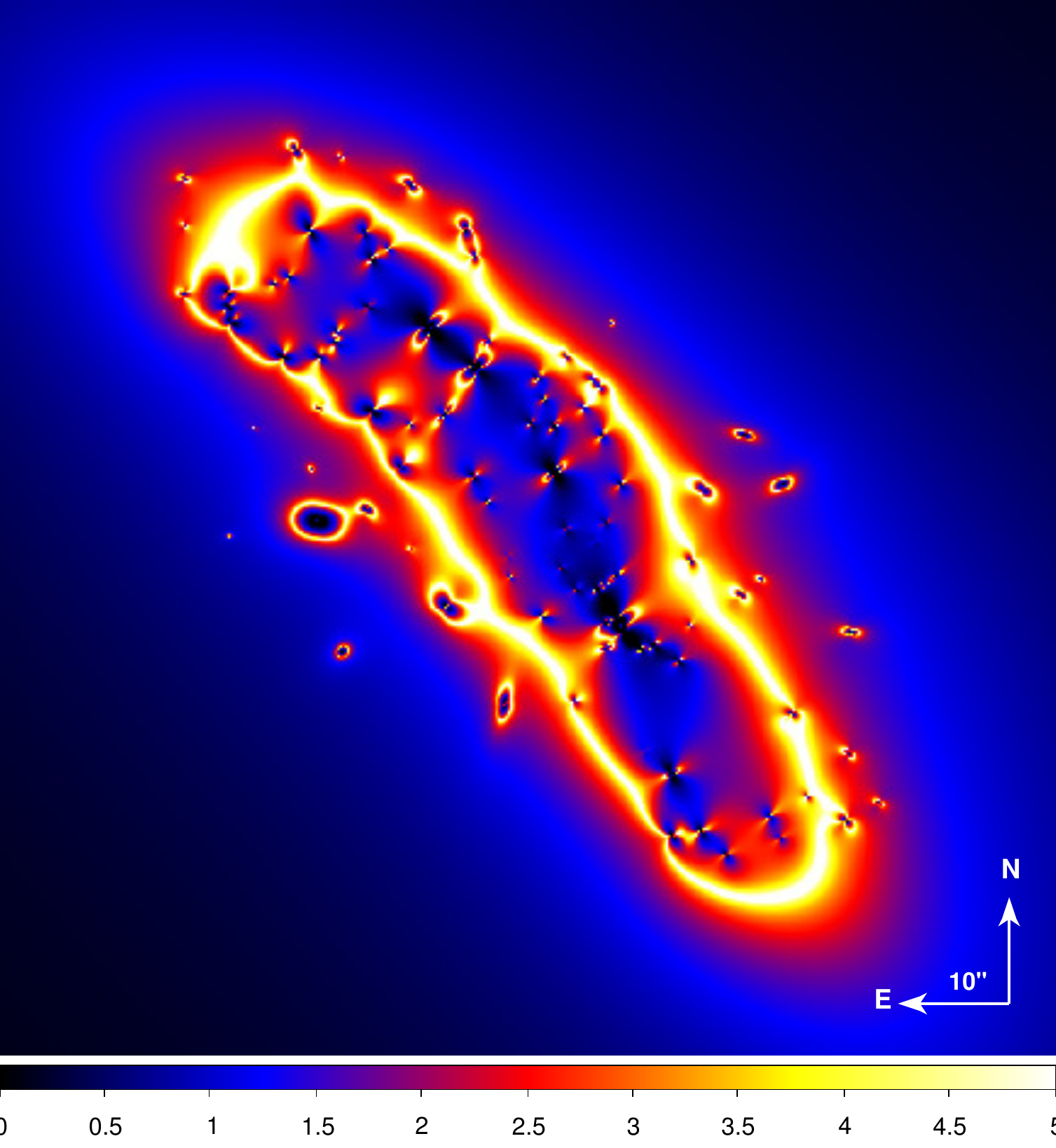}\ 
\hspace*{3mm}\includegraphics[width=5.85cm,angle=0.0]{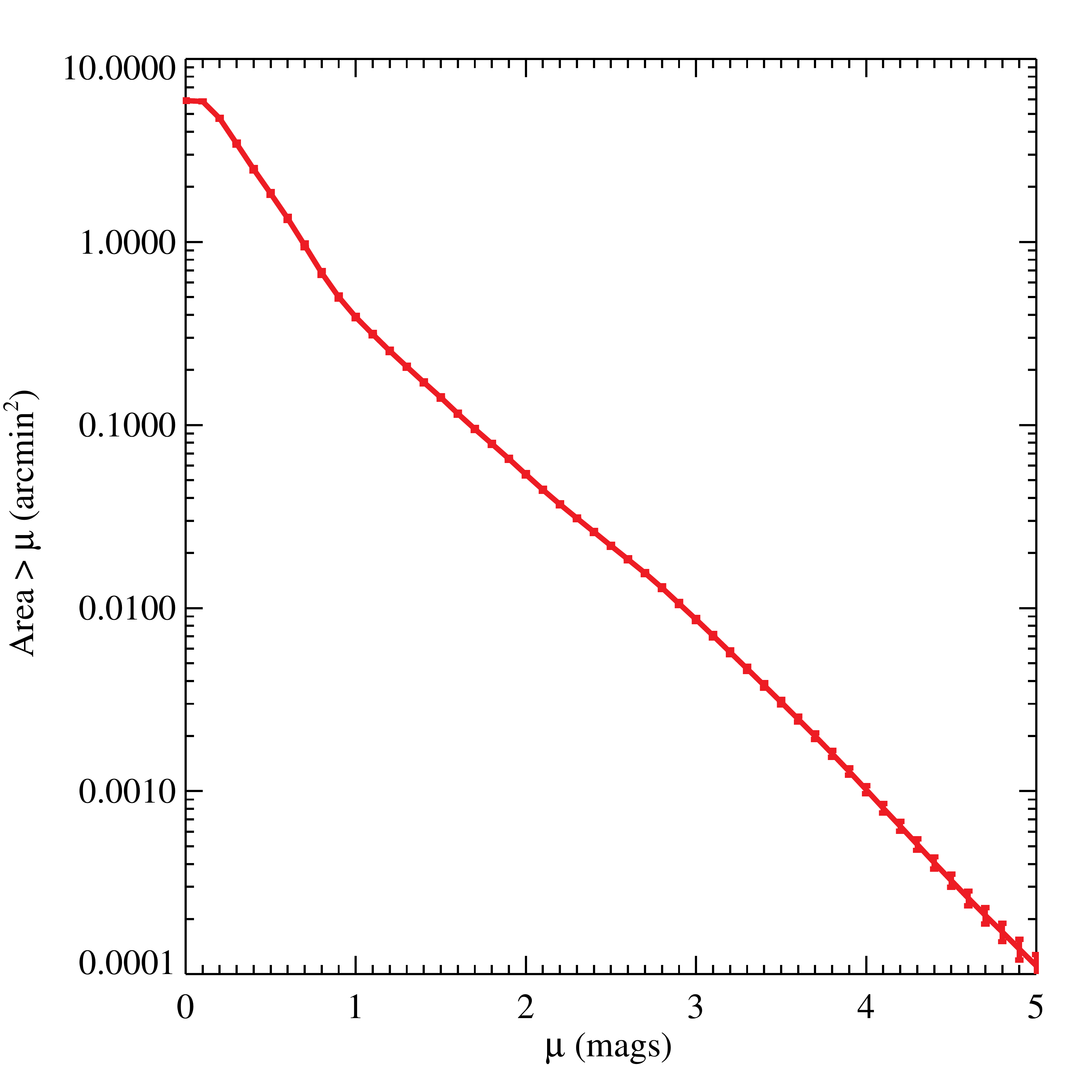}\ 
\includegraphics[width=5.85cm,angle=0.0]{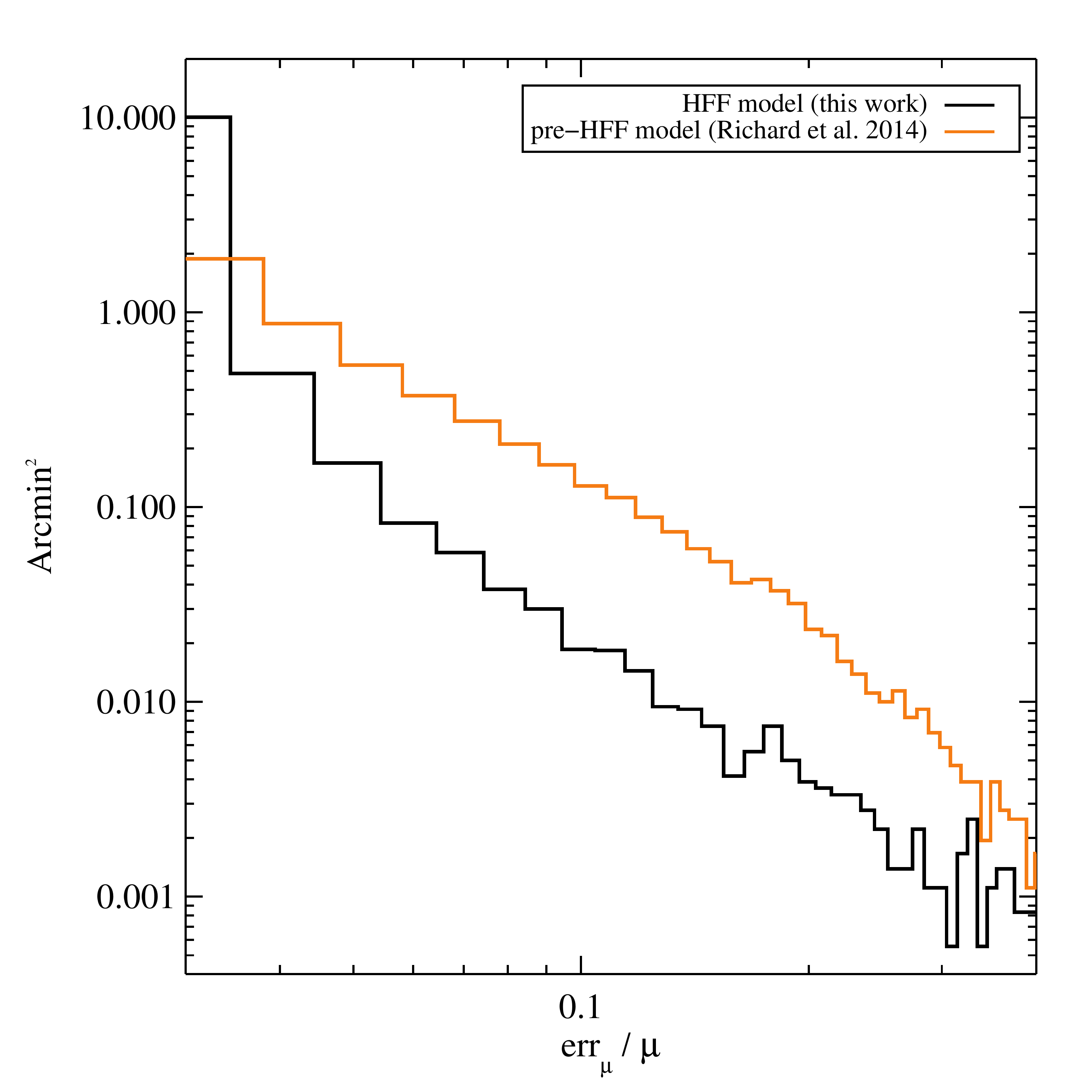}\ 
\\
\caption{\textit{Left panel: }Magnification map obtained from our HFF lens model for a source at $z_{S}=9$. \textit{Middle panel: }Surface area in the source plane covered by ACS at a magnification above a given threshold $\mu$. \textit{Right panel: }Histograms of the relative magnification errors (in linear units) for the pre-HFF lens model of Richard et al.\ (2014) (orange) and our new mass model (black).
}
\label{diffampli}
\end{figure*}
\end{center}

\input{improvementHFF_submitted}

\section*{Acknowledgments}
This work was supported by the Leverhulme Trust (grant number PLP-2011-003) and Science and Technology Facilities Council (grant number ST/L00075X/1). MJ, ML, and EJ acknowledge the M\'esocentre d'Aix-Marseille Universit\'e (project number: 14b030).  This study also benefited from the facilities offered by CeSAM (CEntre de donn\'eeS Astrophysique de Marseille ({\tt http://lam.oamp.fr/cesam/}). MJ thanks Jethro Ridl for his suggestions. ML acknowledges the Centre National de la Recherche Scientifique (CNRS) for its support. Dark cosmology centre is funded by the Danish National Research Fundation. JR acknowledges support from the ERC starting grant CALENDS and the CIG grant 294074. JPK and HA acknowledge support from the ERC advanced grant LIDA. PN acknowledges support from the National Science Foundation via the grant AST-1044455, AST-1044455, and a theory grant from the Space Telescope Science Institute HST-AR-12144.01-A. RM is supported by the Royal Society.


\bibliographystyle{mn2e}
\bibliography{reference}

\input{table_SL_194sys_f814w_amp.tex}



\label{lastpage}

\end{document}

%% file: abstract_submitted.tex
We present a high-precision mass model of the galaxy cluster MACSJ0416.1--2403, based on a strong-gravitational-lensing analysis of the recently acquired \textit{Hubble Space Telescope Frontier Fields} (HFF) imaging data.  Taking advantage of the unprecedented depth provided by HST/ACS observations in three passbands, we identify 51 new multiply imaged galaxies, quadrupling the previous census and bringing the grand total to 68, comprising 194 individual lensed images.  Having selected a subset of the 57 most securely identified multiply imaged galaxies, we use the {\sc Lenstool} software package to constrain a lens model comprised of two cluster-scale dark-matter halos and 98 galaxy-scale halos.  Our best-fit model predicts image positions with an $RMS$ error of 0.68$\arcsec$, which constitutes an improvement of almost a factor of two over previous, pre-HFF models of this cluster. We find the total projected mass inside a 200~kpc aperture to be $(1.60\pm0.01)\times 10^{14}\ M_\odot$, a measurement that offers a three-fold improvement in precision, reaching the percent level for the first time in any cluster.  Finally, we quantify the increase in precision of the derived gravitational magnification of high-redshift galaxies and find an improvement by a factor of $\sim$2.5 in the statistical uncertainty. Our findings impressively confirm that HFF imaging has indeed opened the domain of high-precision mass measurements for massive clusters of galaxies.

%% file: introduction_submitted.tex
The power of gravitational lensing as a tool for observational cosmology was recognised when \cite{a370arcz} spectroscopically confirmed that the giant luminous arc discovered in images of the galaxy cluster Abell 370 (redshift $z\!=\!0.375$) lay far behind the cluster at $z_{arc}\!=\!0.725$. The bending of light from distant galaxies by foreground clusters allows astronomers to  $i)$ directly measure the total (dark and baryonic) matter distribution, $ii)$ image very distant galaxies using galaxy clusters as {`cosmic telescopes'}, and $iii)$ constrain the geometry of the Universe (for reviews, see e.g.\ \citealt{2010RPPh...73h6901M} and \citealt{KN11}). The most massive lenses will produce magnified and highly distorted images of background galaxies, often in multiple-image sets. Strong-lensing analyses of high-quality imaging data in which many ($>$10) such multiply imaged sources are visible enable the most direct and accurate mapping of mass in cluster cores \citep[e.g.][]{bradac06, bradac08a, jullo07, jullo09, coe10}.

The unparalleled power of the \textit{Hubble Space Telescope} (HST) has transformed this field in recent decades. High angular resolution and multi-colour imaging allow the robust and efficient identification of multiple-image systems, as demonstrated in many in-depth studies. For example, using the \textit{Advanced Camera for Survey} (ACS) onboard HST, \cite{broadhurst05a} discovered 30 strongly lensed multiple-image systems behind the massive galaxy cluster Abell 1689 ($z{=}0.183$).  The accuracy of the resulting mass map was further increased by \citet[][]{limousin07b} whose analysis was based on a total of 42 multiply imaged systems, 24 of which were spectroscopically confirmed.

We here present the results of our strong-lensing analysis of a more distant massive cluster. MACSJ0416.1--2403 ($z{=}0.397$; hereafter MACSJ0416) was discovered by the MAssive Cluster Survey  \citep[MACS;][]{ebeling10} and is classified as a merging system based on its double-peaked X-ray surface brightness distribution \citep{ME12}. Because of its large Einstein radius, MACSJ0416 was selected as one of the five ``high-magnification'' clusters in the Cluster Lensing And Supernova survey with Hubble \citep[CLASH:][]{postman12}, resulting in HST imaging in 16 bands from the UV to the near-IR regime, with a typical depth of 1 orbit per passband.  As expected for a highly elongated mass distribution typical of merging clusters, many multiple-image systems were immediately apparent. The first detailed mass model of the system was based on these data and presented by \cite{zitrin13a}. 

The cluster was selected as one of six targets for the \textit{Hubble Frontier Fields}\footnote{http://www.stsci.edu/hst/campaigns/frontier-fields/} (HFF) project, started by the Space Telescope Science Institute in 2013 and aiming to harness the gravitational magnification of massive cluster lenses to probe the distant Universe to unprecedented depth. Using Director's Discretionary Time, the HFF program will observe each cluster for 140 HST orbits, split between three filters on ACS and four on WFC3 (\textit{Wide Field Camera 3}), to reach a depth unprecedented for cluster studies of $mag_{\mathrm{AB}}\!\sim\!29$ in all 7 passbands. Mass models\footnote{http://archive.stsci.edu/prepds/frontier/lensmodels/} of all six HFF cluster lenses were derived from pre-HFF data to provide the community with accurate mass models prior to the arrival of this historical data set \citep[see in particular][]{johnson14,coe14,richard14}.

In this $Letter$ we present results from the first deep ACS observations conducted as part of the HFF initiative and describe the discovery of 51 new multiple-image systems  in HFF images of MACSJ0416 that enabled the first high-precision mass reconstruction of any cluster using nearly 200 multiple images.  We adopt the $\Lambda$CDM concordance cosmology with $\Omega_{m} = 0.3$, $\Omega_{\Lambda} = 0.7$, and a Hubble constant $H_0 = 70$~km$\,$s$^{-1}\,$Mpc$^{-1}$. Magnitudes are quoted in the AB system.

%% file: HFFdata_submitted.tex
The HFF observations of MACSJ0416 (ID: 13496) were obtained with ACS between January 5$^{th}$ and February 9$^{th}$ 2014, in the three filters F435W, F606W, and F814W, for total integration times corresponding to 20, 12, and 48 orbits, respectively.  We applied basic data-reduction procedures using {\tt HSTCAL} and the most recent calibration files. Individual frames were co-added using {\tt Astrodrizzle} after registration to a common ACS reference image using {\tt Tweakreg}. After an iterative process, we achieve an alignment accuracy of 0.1 pixel. Our final stacked images have a pixel size of 0.03\arcsec.

%% file: SLmethod_submitted.tex
We here provide only a brief synopsis of our method which has already been described in detail elsewhere \citep[see, \emph{e.g.}][]{kneib96,smith05,verdugo11,richard11b}.  Our mass model is primarily composed of large-scale dark-matter haloes, whose individual masses are larger than that of a typical galaxy group (of order of 10$^{14}$\,M$_{\sun}$ within 50$\arcsec$), but also takes into account mass perturbations associated with individual cluster members, usually large elliptical galaxies.  As in our previous work, we model all mass components as dual Pseudo Isothermal Elliptical Mass Distributions \citep[dPIEMD,][]{limousin05,eliasdottir07}, characterised by a velocity dispersion $\sigma$, a core radius $r_{\rm core}$, and a scale radius $r_s$.

For mass perturbations associated with individual cluster galaxies, we fix the geometrical dPIE parameters (centre, ellipticity, and position angle) to the values measured from the cluster light distribution \citep[see, \emph{e.g.}][]{kneib96,limousin07b,richard10a}, and use empirical scaling relations (without any scatter) to relate the dynamical dPIE parameters (velocity dispersion and scale radius) to the galaxies' observed luminosity \citep[][]{richard14}.  For an $L_{\ast}$ galaxy, we optimise the velocity dispersion between 100 and 250 km\,s$^{-1}$, and force the scale radius to less than 70 kpc to account for the tidal stripping of galactic dark-matter haloes \citep{limousin07a,limousin09a,natarajan09,wetzel10}.

%% file: SLconstraints_submitted.tex
The secure identification of multiple-image systems is key to building a robust model of the lensing mass distribution. The first strong-lensing analysis of MACS0416  identified 70 images of 23 background sources in the redshift range $0.7\!<\!z\!<\!6$ \citep{zitrin13a}; however, only the 13 most secure systems consisting of 34 individual images were used in the optimisation of the mass model. A combined weak- and strong-lensing analysis based on the same pre-HFF data \citep{richard14} extended the set of secure identifications to 17 multiple-image systems comprising 47 images, by evolving the lens model over several iterations. Nine of these multiple-image systems are spectroscopically confirmed; their spectroscopic redshifts, which range from 1.8925 to 3.2226, are listed in Table~\ref{multipletable}.

For the present study, we scrutinised the new, deep HFF ACS images, using the predictive power of the \citet{richard14} model to find an even larger set of multiple images. To this end, we computed the cluster's gravitational lensing deflection field from the image plane to the source plane, on a grid with a spacing of 0.2 arcsec/pixel. Since the transformation scales with redshift as described by the $D_{LS}/D_{OS}$  distance ratio, the transformation needs to be computed only once. We also determined the critical region at redshift $z=7$ as the area within which to search for multiple images in the ACS data.  A thorough visual inspection of all faint galaxy images in this region, combined with an extensive search for plausible counter images, revealed 68 multiple-imaged systems, comprising 194 individual images (Fig.~\ref{multiples} and Table~\ref{multipletable}).
Table~\ref{multipletable} gives the coordinates, as well as the redshifts (predicted by our model, $z_{model}$, or spectroscopic, $z_{spec}$, if available), the F814W-band magnitudes, $mag_{F814W}$, and their magnification (measured with our best-fit mass model). The magnitudes were measured using \textsc{Sextractor} \cite{BA96}. For some of the images, we could not make reliable measurements due to their proximity to much brighter objects.

For the modelling of the cluster lens, described in detail in the following section, we adopt a conservative approach and use only the 57 most securely identified systems comprising 149 individual images; we propose the remaining identifications as candidate multiple-image systems. 
We consider a system secure if it meets all of the following criteria: the different images have (1) similar colors, (2) show morphological similarities (for resolved images), and finally  (3) a sensible geometrical configuration.
Note that, although the total number of multiple-image sets used in the optimisation has increased by more than a factor of three compared to \citet{richard14}, the area within which they are located has not changed significantly.  As a result, our improved mass model does not extend to much larger radii but dramatically improves the accuracy of the lens mode in the core region of maximal magnification.

%% file: SLmass_submitted.tex
The distribution of cluster galaxies provides a starting point for the modelling process.
In MACS\,J0416, the distribution of light from the cluster ellipticals is elongated along the North-East/South-West direction, with two cD-type galaxies dominating the light budget.
Our initial model thus places one cluster-scale dark-matter halo at the location of each of the two cD-type galaxies that mark the centres of the overall large-scale distribution of light from all cluster galaxies. During the optimisation process, the position of these large-scale halos is allowed to vary within 20$\arcsec$ of the associated light peak. In addition, we limit the ellipticity, defined as $e=(a^2+b^2)/(a^2-b^2)$, to values below 0.7,  while the core radius and the velocity dispersion are allowed to vary between 1$\arcsec$ and 30$\arcsec$, and 600 and 3\,000 km\,s$^{-1}$, respectively. The scale radius, by contrast, is fixed at 1\,000\,kpc, since strong-lensing data alone do not probe the mass distribution on such large scales. In addition to the two cluster-scale dark-matter halos, we also include perturbations by 98 probable cluster members, by associating a galaxy-scale halo to each of them.
Using the set of multiply imaged galaxies described in Sect.~\ref{SLanalysis} and shown in Fig.~\ref{multiples}, we optimise the free parameters of this mass model using the publicly available \textsc{Lenstool} software\footnote{http://projects.lam.fr/repos/lenstool/wiki}.

The unprecedented number of multiple-image systems detected in the HFF observations of MACSJ0416 poses a technical challenge for the ensuing optimisation process.  Not only are not all individual images equally robustly identified, the sheer number of constraints alone proved computationally taxing. Indeed, in order to allow the optimisation of the mass model in the \emph{image plane} with the current version of \textsc{Lenstool} and the computing resources available to us, we had to use a \textsc{rate} parameter \citep[see][]{jullo07} of 0.4  when considering the full set of 57 multiple-image systems. For reference, we usually parametrize the MCMC convergence speed with $\textsc{rate} = 0.1$. By using a considerably larger rate value here, the multi-dimensional parameter space may not be fully sampled, which increases the risk of us missing the best-fit region. 

The best-fit model optimised in the \emph{image plane} predicts image positions that agree with the observed positions to within an RMS of 0.68$\arcsec$. This value is remarkable.  For Abell 1689, the cluster with the previously most tightly constrained mass distribution to date, \cite{broadhurst05b} quote an RMS value of 3.2$\arcsec$, \cite{halkola06} quote 2.7$\arcsec$, and, \cite{limousin07b} quote an RMS value of 2.87$\arcsec$. All these models, as well as ours, are based on the same \emph{a priori} that light traces mass. The RMS value reached by us here for MACSJ0416 thus represents an improvement of a factor of 4 over the residual positional uncertainty of the previously best constrained lensing mass reconstruction.  Using the pre-HFF model of MACSJ0416 as a reference,  the relevant RMS values range from  1.37$\arcsec$ to 1.89$\arcsec$ depending on the model used \citep{zitrin13a}, a factor of 2 larger than the value reached by our high-precision model. 
The parameters describing our best-fit mass model are listed in Table~\ref{tableres}; contours of its mass distribution are shown in Fig.~\ref{multiples}.

To check the robustness of our model we performed the optimisation of the 68 multiple-image systems also in the \emph{source plane}, using our standard value of  \textsc{rate}=0.1. The resulting best-fit model parameters are fully consistent with those derived in our \emph{image-plane} optimisation and are listed in Table~\ref{tableres}. This agreement strongly suggests that the \emph{image-plane} optimisation has indeed converged and instills confidence in the identification of the additional multiple-image systems.  In addition, this second optimisation allowed us to estimate redshifts for all 68 multiple-image systems using the best lens model; we list these redshifts in Table~\ref{multipletable}.

In order to test our initial assumption of a bimodal mass distribution inspired by the large-scale distribution of cluster light, we also investigated a more complex model by associating an additional mass concentration with the bright cluster galaxy located between images 31.2 and 33.1 (Fig.~\ref{multiples}). Given that the resulting RMS of this alternative model with additional free parameters is slightly higher (RMS$=$0.86), we conclude that a third large-scale mass component is not required and not supported by the current observational constraints.

Since the core radii of both cluster-scale halos are large, we can assume that the centre of each of these halos coincides within the error bars with its associated light peak. In order to integrate the mass map within annuli, we choose a centre at $\alpha\!=\!64.0364$, $\delta\!=\!-24.0718$, such that a circle of radius 60$\arcsec$ ($320 h^{-1}$kpc) centered on this point
encompasses all multiple images (Fig.~\ref{multiples}). The two-dimensional (cylindrical) mass within this radius is then $M(<320 \,{\rm h}^{-1}$kpc$) = (3.26\pm0.03) \times 10^{14}$\,M$_{\odot}$.

%% file: modele_SL_submitted.tex

\begin{table}
\begin{center}
\begin{tabular}[h!]{cccc}
\hline
\hline
\noalign{\smallskip}
Clump  & \#1 & \#2 & L$^*$ elliptical galaxy \\
\hline
$\Delta$ \textsc{ra}  & -4.5$^{+0.7}_{-0.6}$  &  24.5$^{+0.5}_{-0.4}$ & --  \\
$\Delta$ \textsc{dec} & 1.5 $^{+0.5}_{-0.6}$  & -44.5$^{+0.6}_{-0.8}$ &  --  \\
$e$ & 0.7 $\pm$0.02  & 0.7$\pm$0.02 & -- \\
$\theta$ & 58.0$^{+0.7}_{-1.2}$  & 37.4$\pm$0.4 & -- \\
r$_{\mathrm{core}}$ (\footnotesize{kpc}) & 77.8$^{+4.1}_{-4.6}$  & 103.3$\pm$4.7  & [0.15] \\
r$_{\mathrm{cut}}$ (\footnotesize{kpc}) & [1000] & [1000] & 29.5$^{+7.4}_{-4.3}$ \\
$\sigma$ (\footnotesize{km\,s$^{-1}$}) &  779$^{+22}_{-20}$ & 955$^{+17}_{-22}$ & 147.9$\pm$ 6.2 \\
\noalign{\smallskip}
\hline
\hline
\end{tabular}
\caption{Best-fit PIEMD parameters for the two large-scale dark-matter halos. 
Coordinates are quoted in arcseconds with respect to $\alpha=64.0381013, \delta=-24.0674860$.
Error bars correspond to the $1\sigma$ confidence level. Parameters in brackets are not optimised.
The reference magnitude for scaling relations is $mag_{\rm{F814W}} = 19.8$.
}
\label{tableres}
\end{center}
\end{table}

%% file: improvementHFF_submitted.tex
The first strong-lensing analysis of MACSJ\,0416 \citep{zitrin13a}, based on pre-HFF data,  estimated the RMS error on predicted image positions as 1.89$\arcsec$ and 1.37$\arcsec$ for mass models parametrised using eGaussian or eNFW profiles respectively, and found a total cluster mass within the effective Einstein radius for a source at $z_{S}=1.896$ of $M(R<145~{\rm kpc})=(1.25\pm0.09)\times 10^{14}$ M$_{\odot}$.  From our current best-fit HFF mass model we derive a slightly lower, but much more precise value of $M(R<145~{\rm kpc})=(1.052\pm0.006)\times 10^{14}$ M$_{\odot}$, an order-of-magnitude improvement in the mass uncertainty and the first time that a cluster mass has been measured to a precision of less than one percent.  Similarly, the dramatic increase in the number of strong-lensing constraints now available led to a reduction by almost a factor of three for the RMS.
Our study thus achieves one of the HFF mission's primary goals: to obtain mass models of massive cluster lenses at an unprecedented level of precision\footnote{We stress in this context that the precision of cluster lensing models depends strongly on the mass modeling technique used in the analysis. For example, our pre-HFF modeling with \textsc{Lenstool} in \cite{richard14} reaches a precision of $\sim$2\% compared to $\sim$ 7\% for the modeling derived by \cite{zitrin13a} from the same imaging data. On-going analysis of FF simulated data will help identify modeling biases, and validate methods of error estimation.}.

Dramatic increases in precision are evident also from a comparison with the pre-HFF mass model presented by \cite{richard14}.  Using a subset of 30 multiple images, the latter yields a median amplification of $4.65\pm0.60$. For the exact same subset of lensed images, but using our current best-fit HFF mass model, we now measured a median amplification of $3.88\pm0.15$, an improvement in {\bf precision} of a factor of four.  In addition, the average error of the predicted positions of the same set of lensed images decreased from RMS${=}1.17\arcsec$ to RMS${=}0.8\arcsec$\footnote{Since these values depend on the subset of multiple-image systems considered, use of only 30 multiple-image families yields a slightly larger value than that reported in Sect.~\ref{SLmass}.}. 

As for the total cluster mass within the multiple-image region, the model of  \cite{richard14} yields $M (R<200~{\rm kpc}) = (1.63\pm0.03)\times 10^{14}$ M$_\odot$ compared to $M (R<200~{\rm kpc}) = (1.60\pm0.01)\times 10^{14}\ M_\odot$ derived from our current HFF mass model.

To summarise, the advent of the HFF data has led to a significant reduction in the statistical errors of both mass and magnification measurements without any change in the analysis and modelling techniques employed.  For MACSJ0416, the fourfold increase in the number of multiple-image systems identified in HST/ACS data lowered the uncertainty in the total mass and magnification by factors of three and four, respectively, making the cluster mass distribution the most tightly constrained yet.  Fig.~\ref{diffampli} summarises our findings by showing the resulting high-fidelity magnification map from our best-fit model, computed for a source at $z_{S}{=}9$, as well as the surface area in the source plane, $\sigma_{\mu}$, above a given magnification factor, which is directly proportional to the unlensed comoving volume covered at high redshift at this magnification. \cite{wong12} proposed using the area above $\mu=3$ as a metric to quantify the efficiency of the lensing configuration to magnify high-redshift galaxies. Our model yields $\sigma_{\mu} (\mu>3) = 0.26$ arcmin$^{2}$ for MACSJ0416. Finally, we also compare in Fig.~\ref{diffampli} the relative magnification errors for our best-fit model and the pre-HFF model of \cite{richard14} 

Owing to the discovery of 51 new multiple-image sets in the HFF/ACS images of MACSJ\,0416, the system's mass map (whose accuracy depends sensitively on the number of lensing constraints) has now reached a precision of better than 1\% in the cluster core, and the uncertainty in the median magnification has been lowered to 4\%.  The resulting high-precision magnification map of this powerful cluster lens immediately and significantly improves the constraints on the luminosity function of high-redshift galaxies lensed by this system, thereby ushering in the HFF era of lensing-aided precision studies of the distant Universe.

%% file: table_SL_194sys_f814w_amp.tex
\newpage
\begin {table}
\begin{tabular}{cc}
\, & \,\\
\end{tabular}
\caption{ Multiply imaged systems considered in this work.
Asterisks indicate the image identifications in which we are less confident. $^{+}$ Even though we have not confirmed system 4 spectroscopically, we assume that systems 3 and 4 correspond to different sub-structures of the \emph{same} background source. Some of the magnitudes are not quoted because we were facing deblending issues that did not allow us to get reliable measurements. The flux magnification factors come from our best-fit mass model, with errors derived from MCMC sampling.}
\label{multipletable}
\end{table}

\tiny

\begin{center}
\par
\tablefirsthead{\hline          \multicolumn{1}{c}{\textbf{ID}} &
                                \multicolumn{1}{c}{\textbf{R.A.}} &
                                \multicolumn{1}{c}{\textbf{Decl.}} &
                                \multicolumn{1}{c}{\textbf{$z_{\rm spec}$}} &
                                \multicolumn{1}{c}{\textbf{$z_{\rm model}$}} &
                                \multicolumn{1}{c}{\textbf{$F814W$}} &
			                    \multicolumn{1}{l}{\textbf{$\mu$}}
                                \\ \hline }
                                
\tablehead{\hline \multicolumn{7}{l}{\small\sl continued from previous page}\\
                         \hline \multicolumn{1}{c}{\textbf{ID}} &
                                \multicolumn{1}{c}{\textbf{R.A.}} &
                                \multicolumn{1}{c}{\textbf{Decl.}} &
                                \multicolumn{1}{c}{\textbf{$z_{\rm spec}$}} &
                                \multicolumn{1}{c}{\textbf{$z_{\rm model}$}} &
                                \multicolumn{1}{c}{\textbf{$F814W$}} &
			                    \multicolumn{1}{l}{\textbf{$\mu$}}
                                \\ \hline  }
\tabletail{\hline\multicolumn{7}{r}{\small\sl continued on next page}\\\hline}
\tablelasttail{\hline}
\par
\begin{supertabular}{ccccccl}
1.1 & 64.04075 & -24.061592 & 1.896 & -- & 25.2 & 5.1$\pm$0.2 \\
1.2 & 64.043479 & -24.063542 & 1.896 & -- & 24.2 & 18.9$\pm$5.1 \\
1.3 & 64.047354 & -24.068669 & 1.896 & -- & 26.0 & 3.1$\pm$0.1 \\
2.1 & 64.041183 & -24.061881 & 1.8925 & -- & 23.6 & 6.0$\pm$0.3 \\
2.2 & 64.043004 & -24.063036 & 1.8925 & -- & 25.2 & 6.4$\pm$0.5 \\
2.3 & 64.047475 & -24.06885 & 1.8925 & -- & 24.1 & 3.0$\pm$0.1 \\
3.1 & 64.030783 & -24.067117 & 1.9885 & -- & 25.5 & 3.3$\pm$0.1 \\
3.2 & 64.035254 & -24.070981 & 1.9885 & -- & 26.6 & 2.2$\pm$0.1 \\
3.3 & 64.041817 & -24.075711 & 1.9885 & -- & 25.2 & 3.2$\pm$0.1 \\
$^{+}$4.1 & 64.030825 & -24.067225 & -- & 1.9 & 24.3 & 3.4$\pm$0.1 \\
$^{+}$4.2 & 64.035154 & -24.070981 & -- & 1.9 & 22.9 & 2.1$\pm$0.1 \\
$^{+}$4.3 & 64.041879 & -24.075856 & -- & 1.9 & 24.3 & 3.0$\pm$0.1 \\
5.2 & 64.032663 & -24.068669 & -- & 1.6 & 24.6 & 14.6$\pm$1.6 \\
5.3 & 64.033513 & -24.069447 & -- & 1.6 & 23.5 & $>$30 \\
7.1 & 64.0398 & -24.063092 & 2.0854 & -- & 25.0 & 10.8$\pm$1.0 \\
7.2 & 64.040633 & -24.063561 & 2.0854 & -- & 25.1 & 23.3$\pm$5.4 \\
7.3 & 64.047117 & -24.071108 & 2.0854 & -- & 28.0 & 2.6$\pm$0.1 \\
8.1 & 64.036596 & -24.066125 & -- & 2.2 & 25.8 & 25.9$\pm$4.6 \\
8.2 & 64.036833 & -24.066342 & -- & 2.2 & 24.0 & $>$30  \\
9.1 & 64.027025 & -24.078583 & -- & 2.25 & 25.6 & 23.0$\pm$3.0 \\
9.2 & 64.027521 & -24.079106 & -- & 2.25 & 25.5 & $>$30 \\
9.3 & 64.036453 & -24.083973 & -- & 2.25 & 28.0 & 2.5$\pm$0.1 \\
10.1 & 64.026017 & -24.077156 & 2.2982 & -- & 24.7 & 6.5$\pm$0.3 \\
10.2 & 64.028471 & -24.079756 & 2.2982 & -- & 24.9 & 5.2$\pm$0.2 \\
10.3 & 64.036692 & -24.083901 & 2.2982 & -- & 25.6 & 2.4$\pm$0.1 \\
11.1 & 64.039208 & -24.070367 & -- & 1.1 & 24.3 & 21.1$\pm$5.1 \\
11.2 & 64.038317 & -24.069753 & -- & 1.1 & 24.0 & $>$30 \\
11.3 & 64.034259 & -24.066018 & -- & 1.1 & 27.0 & 3.3$\pm$0.1 \\
12.1 & 64.038263 & -24.073696 & -- & 1.8 & 25.0 & $>$30 \\
12.2 & 64.037686 & -24.073294 & -- & 1.8 & 25.8 & $>$30 \\
12.3 & 64.029117 & -24.066742 & -- & 1.7 & -- & 2.4$\pm$0.1 \\
13.1 & 64.027579 & -24.072786 & 3.2226 & -- & 25.2 & 7.5$\pm$0.5 \\
13.2 & 64.032129 & -24.075169 & 3.2226 & -- & 23.8 & 3.1$\pm$0.1 \\
13.3 & 64.040338 & -24.081544 & 3.2226 & -- & 25.5 & 2.2$\pm$0.1 \\
14.1 & 64.026233 & -24.074339 & 2.0531 & -- & 23.4 & 4.1$\pm$0.1 \\
14.2 & 64.031042 & -24.078961 & 2.0531 & -- & 23.6 & 2.4$\pm$0.1 \\
14.3 & 64.035825 & -24.081328 & 2.0531 & -- & 23.2 & 4.1$\pm$0.1 \\
15.1 & 64.02686 & -24.075745 & -- & 2.8 & 26.2 & 7.3$\pm$0.4 \\
15.2 & 64.029438 & -24.078583 & -- & 2.8 & 25.3 & 3.8$\pm$0.1 \\
15.3 & 64.038217 & -24.082993 & -- & 2.8 & 25.9 & 2.4$\pm$0.1 \\
16.1 & 64.024058 & -24.080894 & 1.9644 & -- & 23.9 & 6.2$\pm$0.2 \\
16.2 & 64.028329 & -24.084542 & 1.9644 & -- & 22.4 & 8.7$\pm$0.7 \\
16.3 & 64.031596 & -24.085769 & -- & 1.9 & 24.2 & 3.8$\pm$0.1 \\
17.1 & 64.029875 & -24.086364 & 2.2181 & -- & 23.4 & 10.2$\pm$0.8 \\
17.2 & 64.028608 & -24.085986 & 2.2181 & -- & -- & 6.7$\pm$0.3 \\
17.3 & 64.023329 & -24.081581 & 2.2181 & -- & 24.0 & 5.8$\pm$0.2 \\
18.1 & 64.026075 & -24.084233 & -- & 2.1 & 25.6 & 28.6$\pm$6.1 \\
18.2 & 64.025067 & -24.08335 & -- & 2.1 & 25.4 & 26.0$\pm$3.9 \\
18.3 & 64.0309 & -24.086744 & -- & 2.1 & 27.1 & 3.8$\pm$0.1 \\
23.1 & 64.044546 & -24.0721 & -- & 2.1 & 24.8 & 3.6$\pm$0.1 \\
23.2 & 64.039604 & -24.066631 & -- & 2.1 & 25.2 & 1.4$\pm$0.1 \\
23.3 & 64.034342 & -24.063742 & -- & 2.1 & 25.1 & 3.1$\pm$0.1 \\
24.1 & 64.040915 & -24.062959 & -- & 2.2 & 26.9 & $>$30  \\
24.2 & 64.041066 & -24.063057 & -- & 2.2 & 26.0 & $>$30 \\
$^{*}$24.3 & 64.048893 & -24.070871 & -- & 2.2 & 26.6 & 2.3$\pm$0.1 \\
25.1 & 64.044891 & -24.061068 & -- & 2.9 & 25.4 & 14.4$\pm$2.5 \\
25.2 & 64.045448 & -24.061409 & -- & 2.9 & 25.5 & 6.8$\pm$0.6 \\
$^{*}$25.3 & 64.048254 & -24.064513 & -- & 2.9 & 24.9 & $>$30 \\
$^{*}$25.4 & 64.049697 & -24.066948 & -- & 2.9 & 25.8 & 3.3$\pm$0.1 \\
26.1 & 64.04647 & -24.060393 & -- & 5.9 & 26.2 & $>$30 \\
26.2 & 64.046963 & -24.060793 & -- & 5.9 & 27.3 & $>$30 \\
$^{*}$26.3 & 64.049089 & -24.062876 & -- & 5.9 & 27.6 & $>$30 \\
27.1 & 64.048159 & -24.066959 & -- & 2.2 & 24.3 & 5.0$\pm$0.3 \\
27.2 & 64.047465 & -24.066026 & -- & 2.2 & 23.3 & 18.0$\pm$4.9 \\
27.3 & 64.042226 & -24.060543 & -- & 2.2 & 25.6 & 4.6$\pm$0.2 \\
28.1 & 64.036457 & -24.067026 & -- & 1.0 & 23.9 & 15.4$\pm$2.5 \\
28.2 & 64.03687 & -24.067498 & -- & 1.0 & 24.1 & 12.6$\pm$1.8 \\
28.3 & 64.040923 & -24.071151 & -- & 1.0 & 26.4 & 6.8$\pm$0.3 \\
29.1 & 64.034272 & -24.063032 & -- & 2.4 & 25.4 & 2.8$\pm$0.1 \\
29.2 & 64.040131 & -24.066757 & -- & 2.4 & 24.7 & 4.0$\pm$0.2 \\
29.3 & 64.04461 & -24.071482 & -- & 2.4 & 25.8 & 3.7$\pm$0.1 \\
30.1 & 64.033088 & -24.081806 & -- & 4.5 & 26.9 & $>$30 \\
30.2 & 64.032649 & -24.081546 & -- & 4.5 & 27.3 & 17.5$\pm$2.7 \\
31.1 & 64.023833 & -24.077621 & -- & 1.9 & 26.4 & 3.7$\pm$0.1 \\
31.2 & 64.030507 & -24.082725 & -- & 1.9 & 26.5 & 4.7$\pm$0.2 \\
31.3 & 64.032456 & -24.083821 & -- & 1.9 & 25.6 & 5.3$\pm$0.2 \\
32.1 & 64.02413 & -24.08164 & -- & 1.9 & 26.3 & 7.2$\pm$0.3 \\
32.2 & 64.029591 & -24.085572 & -- & 1.9 & -- & 9.4$\pm$0.6 \\
32.3 & 64.030468 & -24.085895 & -- & 1.9 & 28.2 & 5.1$\pm$0.2 \\
33.1 & 64.028427 & -24.082995 & -- & 5.9 & 24.8 & 6.5$\pm$0.3 \\
33.2 & 64.035052 & -24.085486 & -- & 5.9 & 28.1 & 2.9$\pm$0.1 \\
33.3 & 64.02298 & -24.077275 & -- & 5.9 & 26.5 & 4.6$\pm$0.2 \\
34.1 & 64.029254 & -24.073289 & -- & 3.2 & 26.5 & 11.7$\pm$1.0 \\
34.2 & 64.030798 & -24.07418 & -- & 3.2 & 26.8 & 18.9$\pm$2.5 \\
35.1 & 64.037492 & -24.083636 & -- & 4.0 & 26.5 & 2.6$\pm$0.1 \\
35.2 & 64.029418 & -24.079861 & -- & 4.0 & 28.2 & 2.8$\pm$0.1 \\
35.3 & 64.024937 & -24.075016 & -- & 4.0 & 26.5 & 4.3$\pm$0.2 \\
36.1 & 64.02627 & -24.075507 & -- & 3.4 & 25.8 & 6.4$\pm$0.3 \\
36.2 & 64.03842 & -24.083428 & -- & 3.4 & 26.7 & 2.3$\pm$0.1 \\
36.3 & 64.02938 & -24.0789 & -- & 3.4 & -- & 2.4$\pm$0.1 \\
36.4 & 64.029184 & -24.079041 & -- & 3.4 & -- & 2.0$\pm$0.03 \\
37.1 & 64.033791 & -24.082863 & -- & 3.7 & 26.4 & 6.4$\pm$0.4 \\
37.2 & 64.031419 & -24.081613 & -- & 3.7 & 25.5 & 4.9$\pm$0.2 \\
37.3 & 64.022507 & -24.07431 & -- & 3.7 & 26.8 & 2.9$\pm$0.1 \\
38.1 & 64.033625 & -24.083178 & -- & 3.2 & 27.3 & 5.7$\pm$0.3 \\
38.2 & 64.031255 & -24.081905 & -- & 3.2 & 25.5 & 5.1$\pm$0.3 \\
38.3 & 64.022701 & -24.074589 & -- & 3.2 & 28.0 & 2.9$\pm$0.1 \\
$^{*}$39.1 & 64.037335 & -24.072924 & -- & 1.2 & 22.1 & 26.8$\pm$11.8 \\
$^{*}$39.2 & 64.037731 & -24.073135 & -- & 1.2 & 25.2 & 14.1$\pm$2.2 \\
40.1 & 64.037349 & -24.063062 & -- & 3.4 & 28.8 & 5.5$\pm$0.2 \\
40.2 & 64.040346 & -24.064271 & -- & 3.4 & 28.3 & 4.4$\pm$0.2 \\
$^{*}$40.3 & 64.047642 & -24.07443 & -- & 3.4 & 28.4 & 2.3$\pm$0.1 \\
41.1 & 64.037183 & -24.063073 & -- & 3.4 & 28.4 & 5.2$\pm$0.2 \\
41.2 & 64.040369 & -24.064369 & -- & 3.4 & 27.2 & 3.9$\pm$0.1 \\
$^{*}$41.3 & 64.047605 & -24.074313 & -- & 3.4 & 28.2 & 2.3$\pm$0.1 \\
$^{*}$42.1 & 64.045994 & -24.070768 & -- & 2.6 & 27.1 & 3.3$\pm$0.1 \\
$^{*}$42.2 & 64.042073 & -24.065547 & -- & 2.6 & 25.3 & 2.8$\pm$0.1 \\
$^{*}$42.3 & 64.035786 & -24.061938 & -- & 2.6 & 26.7 & 2.9$\pm$0.1 \\
43.1 & 64.035667 & -24.08205 & -- & 2.5 & 27.0 & 3.8$\pm$0.1 \\
43.2 & 64.031195 & -24.079959 & -- & 2.5 & 25.2 & 3.3$\pm$0.1 \\
43.3 & 64.024425 & -24.073603 & -- & 2.5 & 27.1 & 3.1$\pm$0.1 \\
44.1 & 64.045259 & -24.062757 & -- & 3.7 & 25.1 & 12.3$\pm$1.9 \\
44.2 & 64.041543 & -24.059997 & -- & 3.7 & 25.6 & 7.2$\pm$0.4 \\
$^{*}$44.3 & 64.049237 & -24.068168 & -- & 3.7 & 27.4 & 3.1$\pm$0.1 \\
45.1 & 64.035673 & -24.079918 & -- & 1.9 & 26.4 & 5.4$\pm$0.2\\
45.2 & 64.025766 & -24.072231 & -- & 1.9 & 25.0 & 4.1$\pm$0.1 \\
45.3 & 64.032893 & -24.076993 & -- & 1.9 & -- & 1.7$\pm$0.03 \\
46.1 & 64.038256 & -24.080451 & -- & 2.2 & 28.3 & 2.9$\pm$0.1 \\
46.2 & 64.026402 & -24.072239 & -- & 2.2 & 27.6 & 3.9$\pm$0.1 \\
$^{*}$46.3 & 64.033057 & -24.076204 & -- & 2.2 & -- & 3.9$\pm$0.1 \\
47.1 & 64.026328 & -24.076694 & -- & 3.5 & 25.3 & 12.3$\pm$1.1 \\
47.2 & 64.028329 & -24.078999 & -- & 3.5 & 27.2 & 6.7$\pm$0.3 \\
$^{*}$47.3 & 64.038206 & -24.083719 & -- & 3.5 & 27.8 & 2.3$\pm$0.1 \\
48.1 & 64.035489 & -24.084668 & -- & 4.0 & 26.1 & 2.8$\pm$0.1 \\
48.2 & 64.029244 & -24.081802 & -- & 4.0 & 24.0 & 3.8$\pm$0.1 \\
48.3 & 64.023416 & -24.076122 & -- & 4.0 & 25.3 & 3.8$\pm$0.1 \\
49.1 & 64.033944 & -24.074569 & -- & 4.0 & 29.3 & 2.4$\pm$0.1 \\
49.2 & 64.040175 & -24.079864 & -- & 4.0 & 27.0 & 2.7$\pm$0.1 \\
$^{*}$49.3 & 64.026833 & -24.069967 & -- & 4.0 & 24.9 & 6.6$\pm$0.3 \\
$^{*}$50.1 & 64.03479 & -24.074585 & -- & 3.3 & 28.5 & 3.3$\pm$0.1 \\
$^{*}$50.2 & 64.039683 & -24.078869 & -- & 3.3 & 28.4 & 3.2$\pm$0.1 \\
51.1 & 64.04016 & -24.08029 & -- & 4.0 & 26.3 & 2.5$\pm$0.1 \\
51.2 & 64.033663 & -24.074752 & -- & 4.0 & 26.3 & 2.4$\pm$0.1 \\
51.3 & 64.02662 & -24.070494 & -- & 4.0 & 24.9 & 4.3$\pm$0.1 \\
52.1 & 64.045857 & -24.06583 & -- & 4.5 & 25.9 & 10.5$\pm$1.7 \\
52.2 & 64.047698 & -24.068668 & -- & 4.5 & 27.1 & 3.9$\pm$0.2 \\
52.3 & 64.037724 & -24.059826 & -- & 4.5 & 26.6 & 2.9$\pm$0.1 \\
53.1 & 64.046023 & -24.0688 & -- & 3.0 & 25.9 & 5.1$\pm$0.3 \\
53.2 & 64.044776 & -24.066682 & -- & 3.0 & 24.4 & $>$30 \\
$^{*}$53.3 & 64.036197 & -24.060643 & -- & 3.0 & 26.5 & 2.6$\pm$0.1 \\
54.1 & 64.046789 & -24.071342 & -- & 2.4 & 27.0 & 2.8$\pm$0.1 \\
54.2 & 64.041376 & -24.064519 & -- & 2.4 & 26.1 & 3.3$\pm$0.1 \\
54.3 & 64.037157 & -24.062423 & -- & 2.4 & 26.9 & 3.7$\pm$0.1 \\
$^{*}$55.1 & 64.035233 & -24.064726 & -- & 2.6 & 28.1 & 4.9$\pm$0.2 \\
$^{*}$55.2 & 64.04607 & -24.075174 & -- & 2.6 & 28.3 & 2.3$\pm$0.04 \\
$^{*}$55.3 & 64.038514 & -24.065965 & -- & 2.6 & 25.5 & 5.2$\pm$0.2 \\
56.1 & 64.035676 & -24.083589 & -- & 3.3 & 28.2 & 3.1$\pm$0.1 \\
56.2 & 64.030097 & -24.080924 & -- & 3.3 & 28.3 & 3.1$\pm$0.1 \\
56.3 & 64.023847 & -24.074998 & -- & 3.3 & 28.3 & 3.5$\pm$0.1 \\
57.1 & 64.026224 & -24.076036 & -- & 3.0 & 26.4 & 6.2$\pm$0.3 \\
57.2 & 64.028843 & -24.079126 & -- & 3.0 & 26.7 & 3.5$\pm$0.1 \\
58.1 & 64.025187 & -24.073582 & -- & 3.2 & 27.6 & 3.7$\pm$0.1 \\
58.2 & 64.03773 & -24.08239 & -- & 3.2 & 27.4 & 3.0$\pm$0.1 \\
58.3 & 64.030481 & -24.07922 & -- & 3.2 & 25.3 & 2.2$\pm$0.1 \\
59.1 & 64.035851 & -24.072799 & -- & 2.0 & 27.8 & 8.2$\pm$0.7 \\
59.2 & 64.039936 & -24.075622 & -- & 2.0 & 27.9 & 6.0$\pm$0.3 \\
$^{*}$59.3 & 64.029105 & -24.067658 & -- & 2.0 & 28.2 & 2.9$\pm$0.1 \\
60.1 & 64.026724 & -24.07372 & -- & 4.1 & 27.7 & 5.6$\pm$0.3 \\
60.2 & 64.039708 & -24.082514 & -- & 4.1 & 28.3 & 2.2$\pm$0.1 \\
$^{*}$60.3 & 64.030984 & -24.077181 & -- & 4.1 & -- & 20.7$\pm$0.9 \\
61.1 & 64.026732 & -24.07354 & -- & 4.1 & 27.7 & 5.6$\pm$0.3 \\
61.2 & 64.039768 & -24.08236 & -- & 4.1 & 28.0 & 2.2$\pm$0.1 \\
$^{*}$61.3 & 64.030593 & -24.07776 & -- & 4.1 & 28.8 & 3.7$\pm$0.1 \\
$^{*}$62.1 & 64.026889 & -24.07961 & -- & 3.3 & 25.5 & 24.1$\pm$3.6 \\
$^{*}$62.2 & 64.025993 & -24.078892 & -- & 3.3 & 26.4 & 16.5$\pm$1.7 \\
$^{*}$62.3 & 64.036488 & -24.084935 & -- & 3.3 & 28.2 & 2.4$\pm$0.1 \\
63.1 & 64.025535 & -24.07665 & -- & 3.9 & 26.5 & 6.7$\pm$0.4 \\
63.2 & 64.028147 & -24.079648 & -- & 3.9 & 27.2 & 5.1$\pm$0.2 \\
$^{*}$63.3 & 64.037925 & -24.084479 & -- & 3.9 & 27.8 & 2.3$\pm$0.1 \\
$^{*}$64.1 & 64.0431 & -24.07759 & -- & 2.8 & 27.1 & 2.4$\pm$0.1 \\
$^{*}$64.2 & 64.031139 & -24.067177 & -- & 2.8 & 28.0 & 4.1$\pm$0.1 \\
65.1 & 64.042589 & -24.075532 & -- & 5.0 & 26.8 & 3.6$\pm$0.1 \\
65.2 & 64.028858 & -24.064627 & -- & 5.0 & 26.9 & 2.3$\pm$0.1 \\
$^{*}$65.3 & 64.037768 & -24.071656 & -- & 5.0 & 28.0 & 7.3$\pm$0.4 \\
$^{*}$66.1 & 64.038101 & -24.082315 & -- & 2.4 & 28.8 & 2.6$\pm$0.1 \\
$^{*}$66.2 & 64.026635 & -24.074675 & -- & 2.4 & 27.6 & 5.0$\pm$0.2 \\
$^{*}$67.1 & 64.038075 & -24.082404 & -- & 3.2 & 28.7 & 2.7$\pm$0.1 \\
$^{*}$67.2 & 64.025451 & -24.073651 & -- & 3.2 & 27.7 & 3.9$\pm$0.1 \\
$^{*}$67.3 & 64.030363 & -24.079019 & -- & 3.2 & 26.5 & 2.0$\pm$0.1 \\
68.1 & 64.036098 & -24.073362 & -- & 2.8 & 26.3 & 5.5$\pm$0.4 \\
68.2 & 64.040352 & -24.076481 & -- & 2.8 & 23.7 & 6.4$\pm$0.3 \\
68.3 & 64.028017 & -24.06727 & -- & 2.8 & 26.8 & 2.6$\pm$0.1 \\
69.1 & 64.036256 & -24.074225 & -- & 1.6 & 26.9 & 16.1$\pm$3.1 \\
69.2 & 64.037681 & -24.07526 & -- & 1.6 & 28.2 & 10.6$\pm$1.0 \\
$^{*}$69.3 & 64.028759 & -24.069109 & -- & 1.6 & 28.1 & 3.1$\pm$0.1 \\
70.1 & 64.03836 & -24.072385 & -- & 1.5 & 25.6 & $>$30\\
70.2 & 64.03864 & -24.07252 & -- & 1.5 & 27.2 & 19.4$\pm$3.5 \\
$^{*}$70.3 & 64.0321 & -24.06558 & -- & 1.5 & 27.9 & 3.0$\pm$0.1 \\
$^{*}$71.1 & 64.027865 & -24.077908 & -- & 4.6 & 28.5 & $>$30 \\
$^{*}$71.2 & 64.02741 & -24.077382 & -- & 4.6 & 27.7 & 28.6$\pm$4.5 \\
$^{*}$72.1 & 64.031937 & -24.071316 & -- & 2.6 & 28.2 & 12.7$\pm$1.4 \\
$^{*}$72.2 & 64.030952 & -24.07048 & -- & 2.6 & 28.1 & 14.4$\pm$1.9 \\
$^{*}$73.1 & 64.043712 & -24.062603 & -- & 2.4 & 27.4 & 18.4$\pm$4.5 \\
$^{*}$73.2 & 64.041861 & -24.061243 & -- & 2.4 & 29.3 & 6.2$\pm$0.4 \\
\hline
\end{supertabular}
\end{center}